\documentclass[aps,
prx,
twocolumn,
reprint,
noeprint,
superscriptaddress,
amsmath,
amssymb,
]{revtex4-2}
\usepackage[english]{babel}
\usepackage{amsmath}
\usepackage{graphicx}
\usepackage[colorlinks=true, allcolors=blue]{hyperref}
\usepackage{times}
\usepackage{multirow}
\usepackage{bm}
\usepackage{braket}
\usepackage{listings}%

\begin{document}
\title{Scalable high-fidelity and near-deterministic preparation of large photon-number states}

\author{Mo Xiong}
\thanks{These authors contributed equally to this work.}
\affiliation{College of Physics, Nanjing University of Aeronautics and Astronautics, Nanjing 211106, China}

\author{Jize Han}
\thanks{These authors contributed equally to this work.}
\affiliation{China Mobile (Suzhou) Software Technology Co., Ltd., Suzhou, 215163, China}
\affiliation{College of Physics, Nanjing University of Aeronautics and Astronautics, Nanjing 211106, China}

\author{Chuanzhen Cao}
\affiliation{College of Physics, Nanjing University of Aeronautics and Astronautics, Nanjing 211106, China}

\author{Jinbin Li}
\affiliation{College of Physics, Nanjing University of Aeronautics and Astronautics, Nanjing 211106, China}
\affiliation{Key Laboratory of Aerospace Information Sensing and Physics (NUAA), MIIT, Nanjing 211106, China}

\author{Zhiguo Huang}
\affiliation{China Mobile (Suzhou) Software Technology Co., Ltd., Suzhou, 215163, China}

\author{Ming Xue}
\email[Corresponding author: ]{mxue@nuaa.edu.cn}
\affiliation{College of Physics, Nanjing University of Aeronautics and Astronautics, Nanjing 211106, China}
\affiliation{Key Laboratory of Aerospace Information Sensing and Physics (NUAA), MIIT, Nanjing 211106, China}

 \date{\today}

\begin{abstract}
The scalable preparation of large photon-number (Fock) states is a long-standing frontier
in quantum science, with direct implications for quantum metrology and bosonic
quantum information processing.
Despite substantial progress at small photon numbers, extending state
generation to large photon numbers while maintaining high fidelity and operating
deterministically remains a
significant challenge.
Here we demonstrate a scalable and experimentally accessible control protocol for
generating large photon-number states using only native spin--oscillator
operations.
The protocol alternates Jaynes--Cummings interactions with phase-space
displacements to imprint photon-number--dependent phases and convert them into
selective interference in photon-number space.
It already achieves high preparation fidelity unconditionally, 
while an optional final qubit projection removes residual qubit--field
correlations and further enhances the fidelity.
Conditioned on this final projection, photon-number state preparation with fidelities
exceeding $0.95$ is achieved for photon numbers in the few-hundred regime,
with a success probability exceeding $0.90$, placing the protocol in a
near-deterministic operating regime.
The resulting control sequences remain shallow and are robust against
detuning, control noise, and experimentally relevant dissipation.
Our results establish a practical route to scalable, high-fidelity photon-number state preparation
at large photon numbers and provide a versatile interference-engineering toolbox
for nonclassical bosonic state synthesis.
\end{abstract}

\maketitle

\section{Introduction}
The ability to deterministically prepare highly nonclassical bosonic states is a
central goal in quantum science, with far-reaching implications for quantum
metrology, communication, simulation, and fault-tolerant quantum technologies
\cite{braunstein05rmp,loredo2017boson,wang2017high,fabian19motional,
gottesman01encoding,Walschaers21nongaussian,deng2022observing,
winnel2024deterministicGKP,Puri2021GKP,blais2020quantum,hoshi2025entangling}.
Among these states, photon-number (Fock) states $\ket{N}$—energy eigenstates with a precisely
defined number of excitations—constitute fundamental non-Gaussian resources
\cite{grochowski25optimal,sturges19quantum,couteau2023applications}.
As the excitation number increases, large-$N$ photon-number states exhibit increasingly
fine phase-space interference and enhanced sensitivity to external perturbations,
thereby enabling superior performance in precision sensing~\cite{perarnau2020multimode,deng24quantum,fadel2025quantum} and bosonic quantum
information protocols~\cite{zhong2020quantum,madsen2022quantum}.
At the same time, these very properties render their scalable and deterministic
preparation particularly challenging.

Over the past decades, remarkable progress has been achieved in generating
single- and few-excitation photon-number states across a wide range of experimental
platforms \cite{o2010quantum,aharonovich2016singelphotonemi,premaratne2017microwave,wang17converting,migdall2013single,uppu2020scalable,
teja25quantum,rahman25genuine}.
However, extending such capabilities to large photon numbers remains elusive~\cite{hofheinz2008generation,Hofheinz2008measurement,hofheinz2009synthesizing,Bertet20direct,branczyk2010optimized,tiedau2019scalability,
zhou2023fast,sayrin2011real,zhou2012field,peaudecerf2013quantum,Uria20Deterministic,sivak2022model}.
Most existing approaches rely on probabilistic processes, including
measurement-based filtering~\cite{deng24quantum,Teja23distillation,zhang24generating},
or exploit bosonic nonlinearities~\cite{fosel2020efficient,Yanagimoto19adiabatic,lingenfelter2021unconditional,rivera2023creating,zhang2025deterministic},
or dissipation engineering~\cite{li24autonomous}.
Deterministic schemes based on sequential population transfer or photon-number
subtraction or addition~\cite{fink2008climbing,sayrin2011real,premaratne2017microwave,lund24substraction,pasharavesh2025multi}, on the other hand, require an increasing number of elementary operations
as $N$ grows, leading to rapid accumulation of control errors and decoherence.
Despite their effectiveness at small scales, all these approaches are ultimately
constrained by fundamental trade-offs between determinism, robustness, and
scalability.

From a broader control perspective, these difficulties reflect an inherent
trade-off between available control resources and the complexity of the required
control protocols, which becomes increasingly restrictive as the target
photon-number grows.
These challenges point to a more general obstacle: the difficulty of engineering
strong and well-controlled photon-number–dependent phases across a large bosonic
Hilbert space~\cite{heeres15cavity,eickbusch2022fast,landgraf24fast,munoz24photonic,delasHeras25improving}.
Despite important advances, achieving deterministic, scalable, and high-fidelity
preparation of large-$N$ photon-number states remains an open problem, and continued
exploration of diverse and complementary approaches is both necessary and
valuable.

Here we formulate large photon-number state preparation as a task of engineering
interference through time-structured, photon-number--dependent dynamics, rather
than relying on sequential population transfer or bosonic nonlinearities.
This perspective reframes the problem from population engineering to coherent
interference engineering in Fock space.
Our approach exploits the intrinsic nonlinear spectrum of the Jaynes--Cummings (JC)
interaction, which naturally generates photon-number–dependent phase accumulation
while remaining natively available across a wide range of experimental platforms,
including cavity and circuit quantum electrodynamics (QED)~\cite{mivehvar2021cavity,blais21circuit}, trapped ions~\cite{wu1997jaynes,leibfried2003quantum}, and other atomic, molecular, and optical (AMO) systems
\cite{ye24essay}.
By combining JC interactions with phase-space displacement
operations, we construct composite control sequences that deterministically
reshape an initial coherent-state distribution through engineered interference
in photon-number space.

We demonstrate the generation of large photon-number states with fidelities exceeding
$0.95$ up to $N\simeq 200$, with success probabilities above $0.90$ upon
optional post-selection.
The resulting protocols employ only native spin–oscillator operations, require
shallow control depths, and exhibit robustness against detuning, control
noise, and dissipation.
Our results establish a scalable and experimentally accessible route toward
high-fidelity photon-number state generation, opening new opportunities for
quantum-enhanced sensing and bosonic quantum technologies.

\section{Model and Protocol}
\begin{figure}
    \centering
    \includegraphics[width=1.0\columnwidth]{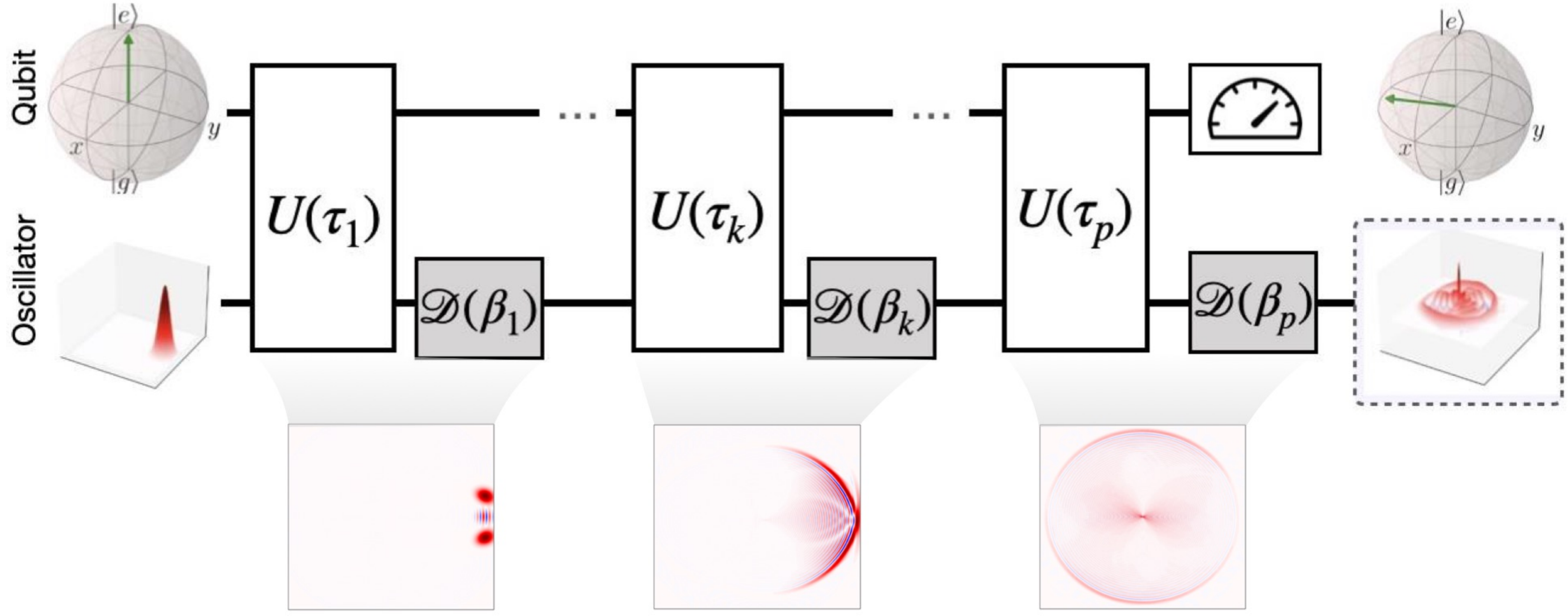}
  \caption{{\bf Multi-pulse JC--displacement protocol for Fock-state generation.}The control sequence consists of $p$ layers, each comprising a resonant
JC evolution $U(\tau_k)$ acting on the joint qubit--oscillator
system, followed by a phase-space displacement ${D}(\beta_k)$ applied to the
bosonic mode.
The qubit is initialized in the excited state and the oscillator in a coherent
state. Repeated accumulation of photon-number--dependent phases arising from the
nonlinear JC spectrum, combined with displacement-induced
interference, progressively reshapes the oscillator state in phase space.
The second row shows representative Wigner-function snapshots at different
layers of the protocol, illustrating the emergence of nonclassical interference
fringes characteristic of large photon-number states.
A final projective measurement on the qubit can be used to disentangle the
qubit from the oscillator.}
\label{fig:model}
\end{figure}

\begin{figure*}[t]
\centering
\includegraphics[width=1.\textwidth]{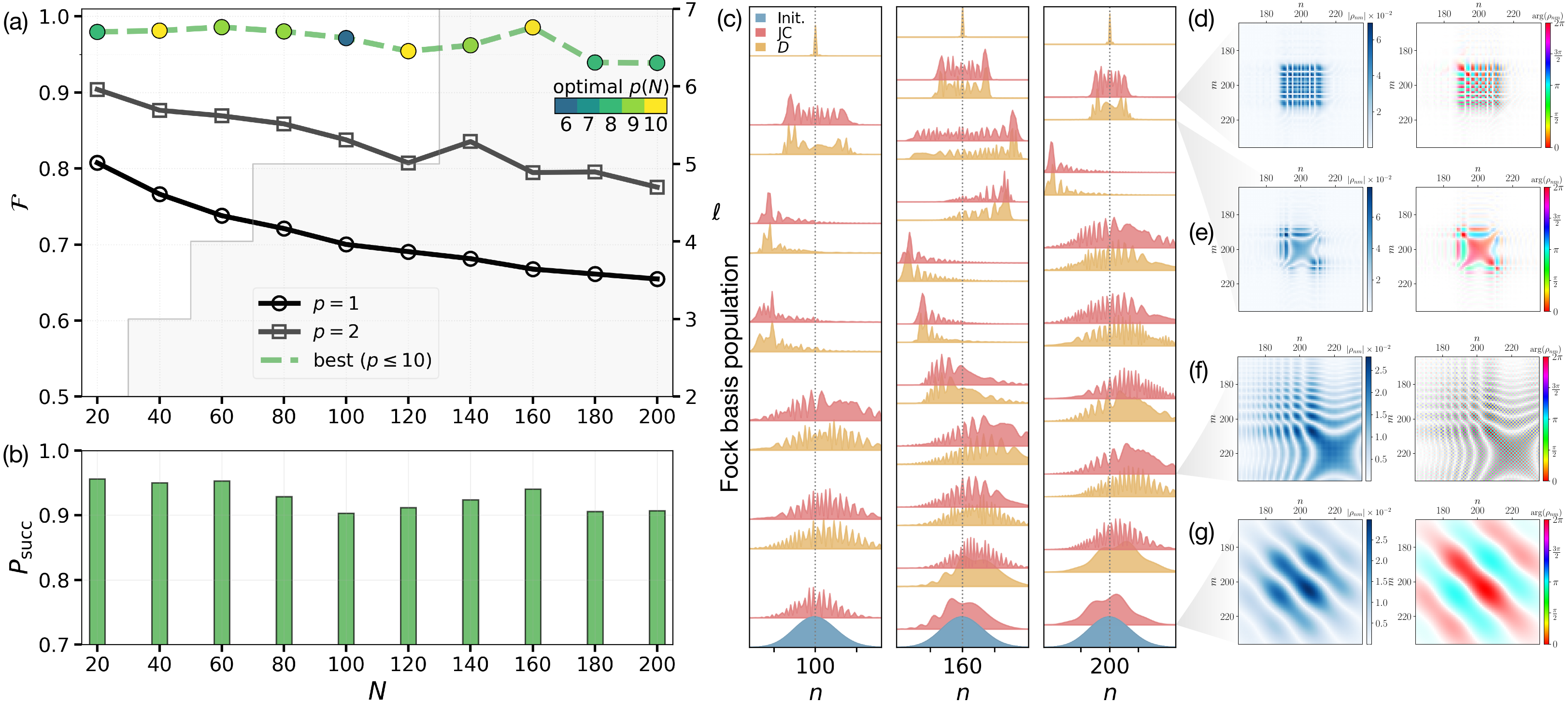}
\caption{
\textbf{Multi-pulse JC–displacement protocol for large-$N$ photon-number state generation.}
(a) Target-Fock fidelity $\mathcal{F}$ as a function of photon-number $N$ for $p=1$, $p=2$, and the optimized multi-pulse protocol ($p\le10$). Colored markers indicate the optimal pulse number $p(N)$ for each $N$. The grey staircase in the background shows the corresponding optimal interaction index $\ell$.
(b) Post-selection success probability $P_{\mathrm{succ}}$ for the optimized protocol, demonstrating that high fidelity is achieved without sacrificing projection efficiency.
(c) Waterfall plots of the photon-number basis populations during the sequential JC interaction and displacement operations (bottom to top) for representative $N=100,160,200$. Each layer shows the population immediately after a JC step (red) and after the subsequent displacement $\hat{D}$ (gold), with the initial Poisson distribution shown in blue. The protocol progressively reshapes the broad coherent-state distribution into a sharply localized photon-number component (vertical dashed line
marks the target $n=N$).
(d)–(g) Density-matrix snapshots illustrating the evolution of magnitude $|\rho_{nm}|$ (left) and phase $\arg(\rho_{nm})$ (right). The multi-pulse sequence first imprints nonlinear number-dependent phases via JC dynamics and subsequently converts them into constructive interference through displacement, resulting in coherent focusing along the target photon-number diagonal. The final state exhibits a dominant diagonal element with suppressed off-diagonal weight, consistent with high-fidelity photon-number state preparation.}
\label{fig:result}
\end{figure*}

\subsection{Composite-pulse protocol}
We consider a spin–boson interaction in which a two-level system couples to a
single-mode quantized field. This standard setting, as described by the JC Hamiltonian, captures the essential physics of light–matter
interaction across cavity QED, circuit QED, and related platforms. 
In the rotating frame ($\hbar=1$), the Hamiltonian reads
\begin{equation}
\hat{H} = -\Delta\hat{a}^\dagger\hat{a}
+\Omega (\hat{a}\hat{\sigma}_++\hat{a}^\dagger\hat{\sigma}_-),
\label{eq:H}
\end{equation}
where $\Delta = \omega_q - \omega_c$ denotes the detuning between the qubit and cavity frequencies ($\Delta=0$ hereafter for the resonant case), and $\Omega$ is the coupling strength.
The operators $\hat{a}$ ($\hat{a}^\dagger$) annihilate (create) photons in the cavity mode, while $\hat{\sigma}_-$ ($\hat{\sigma}_+$) = $|g\rangle\langle e|$ ($|e\rangle\langle g|$) lower (raise) the qubit between its ground and excited states $|g\rangle$ and $|e\rangle$.

Our objective is to prepare a target Fock state $\rho_N=|N\rangle\langle N|$ of
the bosonic mode. The protocol begins with the cavity initialized in a coherent
state $|\alpha\rangle$ of mean photon-number $|\alpha|^2=N$, which provides a
physically natural and experimentally accessible seed for photon-number state
synthesis. The full initial state is therefore
$|\Psi_0\rangle = |e\rangle|\alpha\rangle$.

As depicted in Fig.~\ref{fig:model}, the control sequence consists of \(p\) composite pulses alternating between JC evolution and phase-space displacement operations. 
The sequence is fully specified by
\(\{\boldsymbol{\tau},\boldsymbol{\beta}\}\), where
\(\boldsymbol{\tau}=(\tau_1,\dots,\tau_p)\) and
\(\boldsymbol{\beta}=(\beta_1,\dots,\beta_p)\).
After the full sequence, the joint atom-field state is
\begin{equation}
|\Psi(\boldsymbol{\tau},\boldsymbol{\beta})\rangle = 
\left[\prod_{k=1}^{p} \hat{D}(\beta_k)\hat{U}(\tau_k)\right] |\Psi_0\rangle,
\label{eq:pulse_seq}
\end{equation}
where $\hat{U}(\tau_k) = e^{-i\hat{H}\tau_k}$ is JC evolution of duration $\tau_k\in\mathbb{R}^+$, and
$\hat{D}(\beta_k)=e^{\beta_k\hat{a}^\dagger-\beta_k^*\hat{a}}$ displaces the field by amplitude $\beta_k\in\mathbb{C}$.

A key structural feature of the protocol is the time-control index ($\ell$),
which specifies the total evolution time
through its relation to the JC quantum revival structure~\cite{GeaBanacloche1990,ROBINETT20041}:
\begin{equation}
\Omega T_{\rm R}^{(\ell)}\approx(2\ell+1)\pi\sqrt{N}.
\end{equation}
For a single-layer circuit ($p=1$), choosing $\ell=1$ reproduces the revival-time
operation of Ref.~\cite{Uria20Deterministic}. 
In the multi-pulse setting, however, the interplay between multiple JC segments
and displacement operations generates a much richer interference landscape.  
Accordingly, the control parameters are defined under the global timing scale
set by $T_{\rm R}^{(\ell)}$, while leaving the internal pulse structure unconstrained.

The cavity state resulting from this sequence may be obtained directly by
tracing out the qubit, or, alternatively, via a projective measurement on the
qubit (cf.~Fig.~\ref{fig:model}).
In the latter case, the qubit is projected onto
$|\psi_q\rangle=\cos{(\varphi_{0}/2)}|g\rangle+\sin{(\varphi_{0}/2)}e^{i\varphi_1}|e\rangle$ with $\bm{\varphi} = \{\varphi_0, \varphi_1\}$,
which removes the remaining atom–field correlations and yields a purer cavity state.
This optional post-selection step removes the residual qubit–field correlations
and yields a purer cavity state, with a success probability close to unity.

High-fidelity preparation is obtained by optimizing over the $3p+2$ control
parameters
\begin{equation}
\bm{\theta} = \{\bm\tau,\bm\beta,\bm{\varphi}\},
\label{eq:theta}
\end{equation}
where $\bm\tau=(\tau_1,\dots,\tau_p)$ and $\bm\beta=(\beta_1,\dots,\beta_p)$
specify the durations and complex displacements of the $p$ composite layers, and
$\bm{\varphi}=\{\varphi_0,\varphi_1\}$ defines the (optional) qubit projection
basis.
The quality of the prepared cavity state $\rho_{\bm{\theta}}$ is quantified by
its Uhlmann--Jozsa fidelity with the target Fock state $\rho_N=|N\rangle\langle
N|$,
\begin{equation}
\mathcal{F}(\bm\theta) =
\left(\mathrm{Tr}\sqrt{\sqrt{\rho_{\bm{\theta}}}\rho_N\sqrt{\rho_{\bm{\theta}}}}\right)^2.
\end{equation}
To avoid repeated reduced-state evaluations during the search, we instead
minimize the proxy loss
\begin{equation}
\mathcal{L}_{\bm{\theta}}
=1-|\langle\Psi(\bm\tau,\bm\beta)|\Psi_{N}(\bm{\varphi})\rangle|^{2},
\label{eq:loss}
\end{equation}
where $|\Psi_N(\bm{\varphi})\rangle = |\psi_q(\bm{\varphi})\rangle\otimes|N\rangle$
denotes the ideal disentangled target state.
This variational formulation naturally generalizes the single-pulse JC approach
to a richer multi-pulse control landscape and enables efficient numerical
optimization of composite sequences.

In practice, the non-convex landscape associated with Eq.~(\ref{eq:loss})
contains many periodic local minima.
Throughout this work, the parameters $\bm\theta$ are obtained using a hybrid
genetic--Adam (GAdam) optimization routine. To accelerate the search and enforce a physically meaningful global timing
scale, the depth-$p$ optimization is typically seeded via transfer
initialization from an optimized single-layer ($p=1$) solution at the same
$(N,\ell)$, which provides an effective starting point for the subsequent
multi-pulse refinement (see Methods for detail).

\subsection{Fidelity and dynamics}
Fig.~\ref{fig:result} benchmarks the our protocol and
highlights its encouraging scaling with photon-number $N$.
As seen in Fig.~\ref{fig:result}(a), increasing the depth from the single-layer
baseline ($p=1$) to $p=2$ already mitigates the fidelity loss with $N$, while the
optimized protocol (best, $p\le 10$) exhibits only weak degradation over the explored range.
The optimized sequences achieve post-selected fidelities $\mathcal{F}\ge 0.95$
for all targets up to $N=160$, and remain only slightly below $0.95$ at
$N=180,200$.
The optimal depth $p(N)$ (colored markers) grows slowly with $N$, while the
interaction-time index $\ell$ (grey staircase) sets a global timing scale.
High fidelity is retained in a near-deterministic regime, with
$P_{\mathrm{succ}}\gtrsim 0.90$ throughout [Fig.~\ref{fig:result}(b)], indicating
that the final projection primarily purifies the cavity state by removing
residual qubit--field correlations.
The mild degradation at the largest $N$ is obtained under a fixed optimization
budget and is therefore conservative.

\begin{figure*}
\centering
\includegraphics[width=1.98\columnwidth]{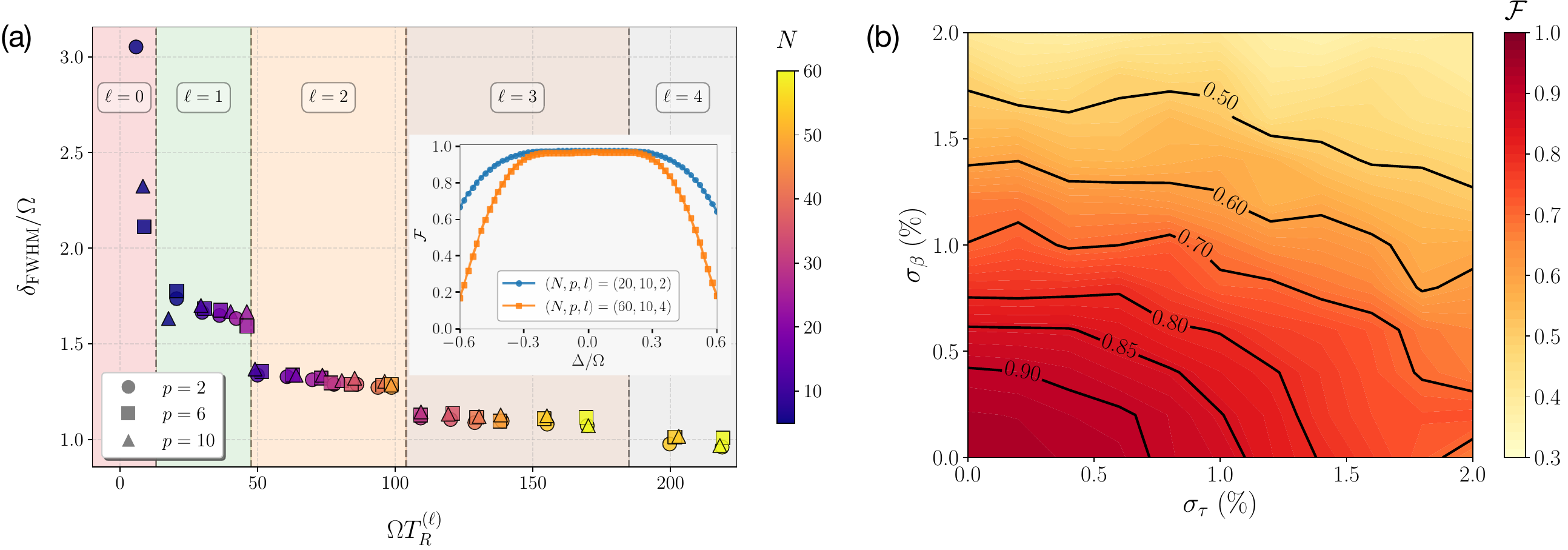}
\caption{
\textbf{Robustness of the multi-pulse protocol against detuning, control noise.}
\textbf{(a)} Detuning tolerance quantified by the full width at half maximum (FWHM) of $\bar{\mathcal{F}}(\Delta/\Omega)$ as a function of the protocol time scale $\Omega T_R^{(\ell)}$ for different $\ell$-families.
Marker shapes denote the circuit depth $p$ and colors indicate the target photon-number $N$.
Inset: representative linecuts of ${\mathcal{F}}(\Delta/\Omega)$ at $p=10$.
\textbf{(b)} Average post-selected fidelity $\bar{\mathcal{F}}$ under timing and displacement errors for $N=100$ ($p\!=\!10,\ell\!=\!4$), averaged over 200 realizations of independent Gaussian perturbations with standard deviations $\sigma_\tau$ and $\sigma_\beta$.
}
\label{fig:robustness}
\end{figure*}

The dynamical reshaping is visualized in the waterfall plots of
Fig.~\ref{fig:result}(c) for representative
$N=100,160,200$.
Starting from the broad Poissonian seed (blue) and proceeding from bottom
to top along the pulse sequence, each JC segment (red) introduces
pronounced oscillatory structure while redistributing population only locally
within the existing photon-number support.
The subsequent displacement (gold) then produces the dominant envelope reshaping
and a systematic concentration toward $n=N$, culminating in a sharply localized final peak.

Each JC segment reshuffles populations only locally in the Fock basis and transfers weight only between adjacent photon-number states and hence preserves a
broad envelope at early stages.
The dominant role of the JC segment is instead to imprint a nonlinear,
number-dependent phase structure onto photon-number space coherences.

The phase snapshots $\arg(\rho_{nm})$ in Figs.~\ref{fig:result}(d)--(g) provide a
direct signature of the nonlinear JC spectrum (right panels).
Because the resonant dressed splittings $E_{n,\pm}=\pm\Omega\sqrt{n+1}$ vary
nonlinearly with photon-number, different photon-number components accumulate different
phases.
In the semiclassical pointer-state picture~\cite{GeaBanacloche1990}, this
nonlinearity can be made explicit by expanding $\sqrt{n}$ for $n$ in the
vicinity of a large target photon-number $N$,
\begin{equation}
\sqrt{n}
=
\sqrt{N}
+\frac{\delta n}{2\sqrt{N}}
-\frac{\delta n^{\,2}}{8N^{3/2}}
+\cdots,
\qquad
\delta n \equiv n-N,
\end{equation}
which separates a leading linear phase gradient from a shear-like quadratic
contribution.
This number dependence first appears as clear band-like phase structures
[Fig.~\ref{fig:result}(g)], and then evolves into visibly curved bands as
higher-order contributions become important [Figs.~\ref{fig:result}(f) and
\ref{fig:result}(e)].
In later layers, repeated alternation of JC evolution (nonlinear phase
accumulation) and displacement (phase-sensitive mixing) progressively organizes
the phase texture, approaching an anti-diagonal alignment of the mirror-symmetric
coherence ridge near $n+m\simeq 2N$ immediately before the final displacement step,
thereby preparing the interference condition for the final focusing.

\subsection{Robustness analysis}
We evaluate the robustness of the protocol against experimentally relevant
imperfections, grouped into (i)~systematic control errors, including qubit--oscillator
detuning and inaccuracies in pulse timing and displacement amplitude; and
(ii)~environment-induced errors from open-system dynamics such as cavity photon
loss and qubit spontaneous emission.

\paragraph*{Detuning robustness.}
We first consider qubit--oscillator detuning
$\Delta=\omega_q-\omega_c$.
The dependence of the post-selected fidelity on detuning is illustrated in the
inset of Fig.~\ref{fig:robustness}(a) for a representative circuit depth
$p=10$.
For small photon numbers ($N=20$), the fidelity remains close to unity
over a broad range of $\Delta/\Omega$, whereas for larger photon numbers
($N=60$) the high-fidelity plateau becomes noticeably narrower, reflecting the
increased sensitivity of high-$N$ photon-number states to detuning-induced phase
accumulation during the JC evolution.
This behavior is quantified in Fig.~\ref{fig:robustness}(a), where the detuning
tolerance is characterized by the full width at half maximum
$\delta_{\mathrm{FWHM}}$ as a function of the total evolution time
$\Omega T^{(\ell)}_{\rm R}$ for different circuit depths $p$.
For small photon numbers or short total evolution times,
$\delta_{\mathrm{FWHM}}$ shows a clear dependence on the number of control
layers: increasing $p$ introduces additional phase structure and control
complexity, which can reduce the effective detuning bandwidth.
In contrast, for larger photon numbers the data for $p=2,6,$ and $10$ collapse
onto a common trend and the extracted $\delta_{\mathrm{FWHM}}$ tends to
saturate, indicating a time-limited regime in which detuning-induced phase
errors are governed primarily by the total accumulated interaction time rather
than by the specific segmentation of the sequence; this saturation is favorable
for high-$N$ operation since the detuning bandwidth does not continue to shrink
rapidly as $N$ increases.
\paragraph*{Pulse errors.}
We next analyze the impact of control inaccuracies in the JC evolution times and
displacement amplitudes.
The implemented control parameters are modeled as
\begin{equation}
\tau_k = \tau_k^{\star}+ \xi_k^{\tau}, \qquad
\beta_k = \beta_k^{\star} + \xi_k^{\beta},
\end{equation}
where the noise variables $\xi_k^{\tau}$ and the real and imaginary parts of
$\xi_k^{\beta}$ are sampled independently from zero-mean Gaussian
distributions with variances $\sigma_\tau^2$ and $\sigma_\beta^2$,
respectively.

Fig.~\ref{fig:robustness}(b) shows the resulting fidelity landscape for the
$N=100$ target with $p=10$ and $\ell=4$,
where each point corresponds to the post-selected fidelity
$\bar{\mathcal{F}}$ averaged over 200 independent realizations of the control
errors.
A broad high-fidelity basin persists around the optimum.
In particular, {sub-percent} timing errors and displacement
miscalibrations preserve a near-ideal performance:
for $(\sigma_\tau,\sigma_\beta)\lesssim(0.5\%,0.5\%)$, we find
$\bar{\mathcal{F}}\gtrsim 0.90$ for the $N=100$ target.
For qubit--oscillator coupling strengths in the range
$\Omega/2\pi \sim \mathrm{kHz}$--$\mathrm{MHz}$, typical of current cavity or circuit QED
platforms, the total evolution time $T_N$ spans from several tens of milliseconds down to the sub-millisecond regime.
Each JC segment therefore has a characteristic duration
$\tau_k \sim T_N/p$, placing it in the sub-millisecond to millisecond range across these platforms.
In practical AMO experiments~\cite{ye24essay}, the evolution time of a gate operation is typically
controlled by an arbitrary waveform generator with sampling rates reaching the gigahertz regime, enabling temporal accuracy at the level of a few tens of nanoseconds.
Comparing these timescales indicates that relative timing errors can be well controlled below the percent level under realistic experimental conditions~\cite{sun2014tracking}.
The displacement amplitudes can be calibrated with reliable experimental control
and stability
~\cite{burkhart21error,eickbusch2022fast,hoshi2025entangling,deng24quantum}.

\paragraph*{Dissipation effects.}
We finally evaluate the performance of the multi-pulse protocol in the presence of
open-system dynamics arising from cavity photon loss and atomic spontaneous
emission. The system evolution is modeled by the Lindblad master equation
$ \dot{\rho}(t) \!=\! -i[\hat{H},\rho(t)]
 + \frac{\kappa}{2}\mathcal{D}[\hat{a}]\,\rho 
 + \frac{\Gamma}{2}\mathcal{D}[\hat{\sigma}_-]\rho,$
where $\kappa$ and $\Gamma$ denote the cavity decay and atomic spontaneous
emission rates, respectively, and
$\mathcal{D}[\hat{o}]\rho=2\hat{o}\rho\hat{o}^\dagger
-\hat{o}^\dagger\hat{o}\rho-\rho\hat{o}^\dagger\hat{o}$
is the Lindblad superoperator. A convenient dimensionless figure of merit that
summarizes the competition between coherent JC dynamics and dissipation is the
single-atom cooperativity $C \equiv {\Omega^{2}}/{\kappa\,\Gamma}$.

A particularly promising direction is provided by atom-array cavity QED, where
two-dimensional ordered atomic arrays in free space form an effective optical
cavity with well-defined cavity-QED parameters.
Recent cavity-array experiments and near-term proposals already operate in the
above-unity regime and indicate feasible pathways toward
$C\sim\mathcal{O}(1)$ or higher~\cite{yan23superradiant,liu23realization,adam26cavityarray}.
For the ${}^{87}\mathrm{Rb}$ D-line, Ref.~\cite{castells25cavity} further
estimates $\Gamma\simeq 2\pi\times 6~\mathrm{MHz}$ and an effective cavity loss
rate $\kappa\simeq 2\pi\times 0.06~\mathrm{MHz}$ in a motion-limited lattice
configuration, while demonstrating that optimized atomic placement can in
principle reach cooperativities as large as $C\gtrsim 10^{4}$.

Importantly, these constraints can be further relaxed by exploiting collective
light--matter coupling in atomic ensembles~\cite{liu23realization,mivehvar2021cavity,
pennetta2022observation,blaha22beyond,castells25cavity,adam26cavityarray}.
In the single-excitation manifold, symmetric collective states enhance the
coupling as $\Omega_{\mathrm{eff}}=\Omega\sqrt{N_{\!\mathrm{atom}}}$~\cite{Fleischhauer02quantum},
thereby shortening the required evolution time and reducing the impact of
dissipation.
Using an experimentally motivated collective cooperativity benchmark
$N_{\!\mathrm{atom}}C\sim 10^{8}$ and and taking the unitary-optimized value
$\mathcal{F}\approx 0.97$ at $N=100$ from Fig.~\ref{fig:result}(a), 
we estimate that post-selected preparation remains feasible with
$\mathcal{F}\gtrsim 0.8$ while maintaining a high-success-probability
operating regime.
Together with the robustness to control errors discussed above, this suggests
that deterministic high-fidelity preparation of photon-number states beyond $N=100$ is within
reach in state-of-the-art AMO cavity-QED platforms.

\section{Conclusion}

In this work, we introduced a multi-pulse JC--displacement protocol
for scalable generation of large photon-number states.
By alternating JC evolutions with phase-space displacements, the
protocol engineers photon-number--dependent phases and converts them into
selective interference in photon-number space, yielding high-fidelity preparation of
photon-number states with $\mathcal{F}\gtrsim 0.95$ up to $N\simeq 200$ using shallow
control depths.
An optional qubit projection further removes residual qubit--field correlations
with near-unity success probability.
In addition, incorporating platform-specific constraints and dissipation into
the optimization loop offers a practical pathway toward hardware-ready pulse
sequences and further scaling to larger photon numbers.

Our results provide an experimentally accessible route to large-$N$ photon-number
state preparation using only native qubit--oscillator operations, applicable to
cavity and circuit QED, trapped ions, and related AMO platforms~\cite{florian14Entangled,Srivastava26entanglement}.
Looking forward, the same interference-engineering approach can be extended to
prepare more general nonclassical bosonic resources, such as photon-number state
superpositions and grid-like states~\cite{Brock25quantum}, and to optimize probe states for quantum-enhanced sensing~\cite{valahu25Quantum-enhanced}.

\paragraph*{Note added.}
During the preparation of this manuscript, we became aware of a closely related
preprint that introduces a similar photon-number space interference approach based on Kerr-induced nonlinear phase modulation for generating macroscopic photon-number states
\cite{li2026scalable,xu2026principlesopticsfockspace}.

\section*{Acknowledgements}
This work was supported by the National Natural Science Foundation of China (No.~12304543), the Quantum Science and Technology - National Science and Technology Major Project (No.~2021ZD0302100), the Natural Science Foundation of Jiangsu Province (Nos.~BK20250404 and BG2025017), the Youth Science and Technology Talent Support Program of Jiangsu Province (Nos.~JSTJ-2024-507, JSTJ-2025-600 and JSTJ-2025-829), the Frontier Technology Research Program of Suzhou (No.~SYG202322), and the Postgraduate Research \& Practice Innovation Program of NUAA (No.~xcxjh20252107).

\appendix
\section{Revivals and phase-space refocusing in the resonant JC model}
\begin{figure}[ht]
    \includegraphics[width=1\columnwidth]{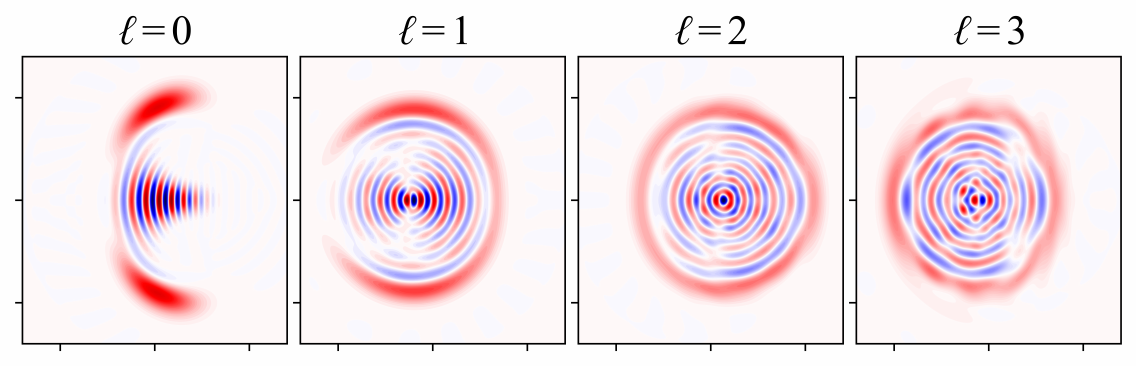}
    \caption{At discrete revival times (labeled by $\ell$), the bosonic state refocuses in
    phase space and becomes Fock-like under the resonant Jaynes--Cummings
    Hamiltonian.}
    \label{fig:time_revival}
\end{figure}

Starting from $\ket{\Psi_0}=\ket{e}\ket{\alpha}$ with
$\ket{\alpha}=e^{-|\alpha|^2/2}\sum_{n=0}^{\infty}\alpha^n/\sqrt{n!}\,\ket{n}$,
the resonant Hamiltonian in Eq.~(\ref{eq:H}) ($\omega_c=\omega_q$) produces the
well-known collapse-and-revival dynamics of the cavity field~\cite{Eberly1980,YOO1985239,GeaBanacloche1990}.
At discrete revival times (indexed by $\ell$), the field refocuses in phase space,
as illustrated in Fig.~\ref{fig:time_revival}.

Following the displacement $\mathcal{ \hat D}(\beta)=e^{\beta\hat{a}^\dagger-\beta^\ast\hat{a}}$,
the qubit-cavity state transforms into:
\begin{eqnarray}
    |\Psi(\beta,t)\rangle &=& \mathcal{ \hat D}(\beta) e^{-i\hat{H}t}|\Psi_0\rangle
      = \sum_{n=0}^{\infty}   |K_n (t)\rangle\otimes |n\rangle,\label{eq:Psi_beta}
\end{eqnarray}
where $ |K_m (t)\rangle $ represents the state of the atomic component. So the density matrix of the qubit-cavity state is: 
\begin{equation}
    \rho_{qc}(\beta, t)=\sum_{m=0}^{\infty}\sum_{m^\prime=0}^{\infty} |K_m (t)\rangle \langle K_{m^\prime}(t)|\otimes |m\rangle \langle m^\prime|.
\end{equation}
$|K_m\rangle$ describes the state of the qubit conditioned on the cavity field being in the photon-number state $|m\rangle$ after displacement by $D(\beta)$.
By tracing out the qubit degrees of freedom, 
one obtains the reduced density operator for the cavity field:
\begin{equation}
    \rho_c (\beta, t ) \equiv {{\rm Tr}_{q}}[\rho(\beta, t)] = \sum_{m=0}^{\infty} \sum_{n=0}^{\infty} P_{m,n}(\beta, t)|m\rangle \langle n|,
\end{equation}
where $P_{m,n}$ denotes the coefficient of the matrix elements $|m\rangle \langle n|$.
Therefore, the occupation probability of the photon-number state $|n\rangle$ is given by:

\begin{align}
    P_{N}(t) \!=\!\!\sum_{m,n=0}^{\infty}\!\!\! c_m
    {c_{n}^{*}} |\langle n |{\hat D}(\beta)|m\rangle|^2
     (s_{m+1}s_{n+1}\!
     +\!\!\frac{\sqrt{m n}}{|\alpha|^2}\!s^\prime_{m}s^\prime_{n}),
     \label{eq:Fmm}
\end{align}
where $s_m = \cos{(\Omega\sqrt{m}t)}$, $s^\prime_m = \sin{(\Omega\sqrt{m}t)}$ 
and $c_m=\alpha^n e^{-|\alpha|^2/2}/\sqrt{n!}$.

The dominant contribution in this expression comes from terms where $m \approx n\approx|\alpha|^2$ ($m, n \gg 1$),
as determined by both the high-frequency oscillating exponential factor and energy considerations. 
photon-number like states, particularly when $m\approx n\approx N$,
are approximately realized at specific time instants given by
\begin{equation}
\Omega T_{\rm R}^{(\ell)}\approx(2\ell+1)\pi\sqrt{N}.
\end{equation}
where $\ell\in\mathbb{N}$.

\begin{figure}[ht]
    \includegraphics[width=.9\columnwidth]{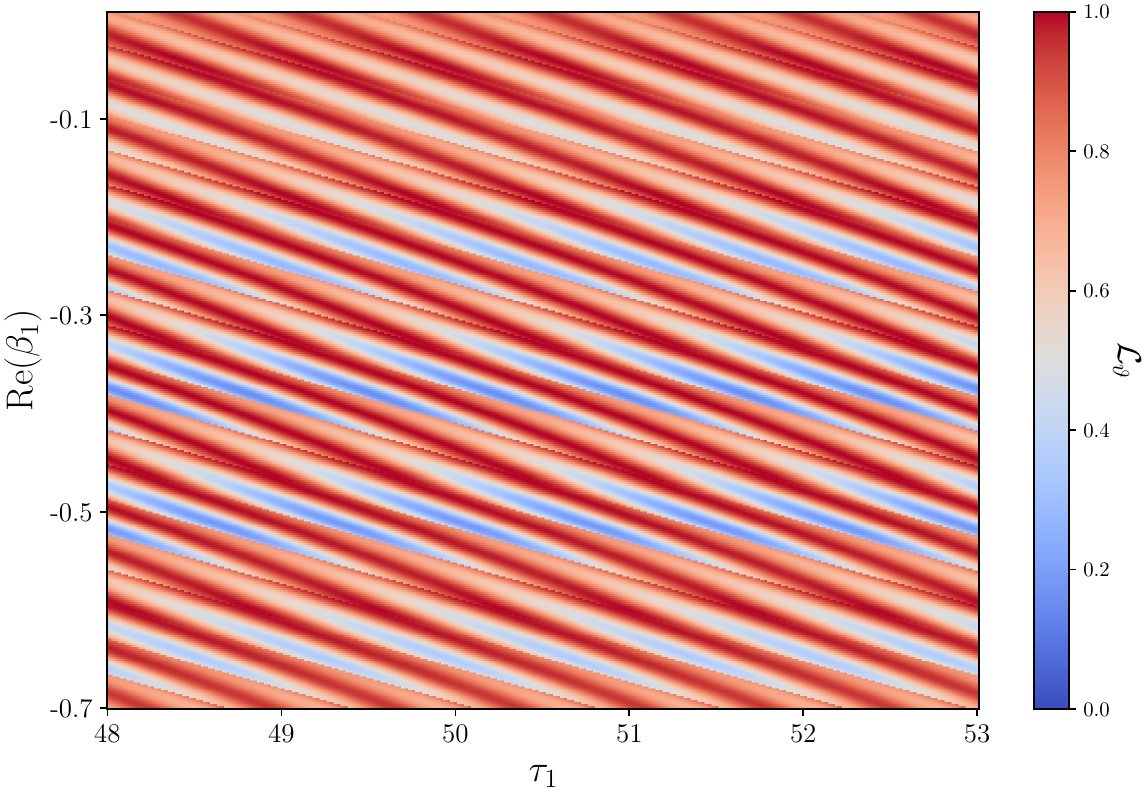}
    \caption{{Non-convex loss landscape in a reduced subspace.}
    Reduced loss $\mathcal{L}_{\bm{\vartheta}}$ after a single JC segment
    followed by one displacement ($p=1$), shown as a function of the interaction time $\tau_1$
    and the displacement quadrature $\mathrm{Re}(\beta_1)$ (with $\mathrm{Im}(\beta_1)$ fixed).
    The quasi-periodic stripe pattern reveals a dense set of competing local extrema, highlighting
    the intrinsic difficulty of the optimization even in this two-parameter slice.}
    \label{fig:nonconvex}
\end{figure}

\begin{figure}[t]
    \centering
    \includegraphics[width=0.99\columnwidth]{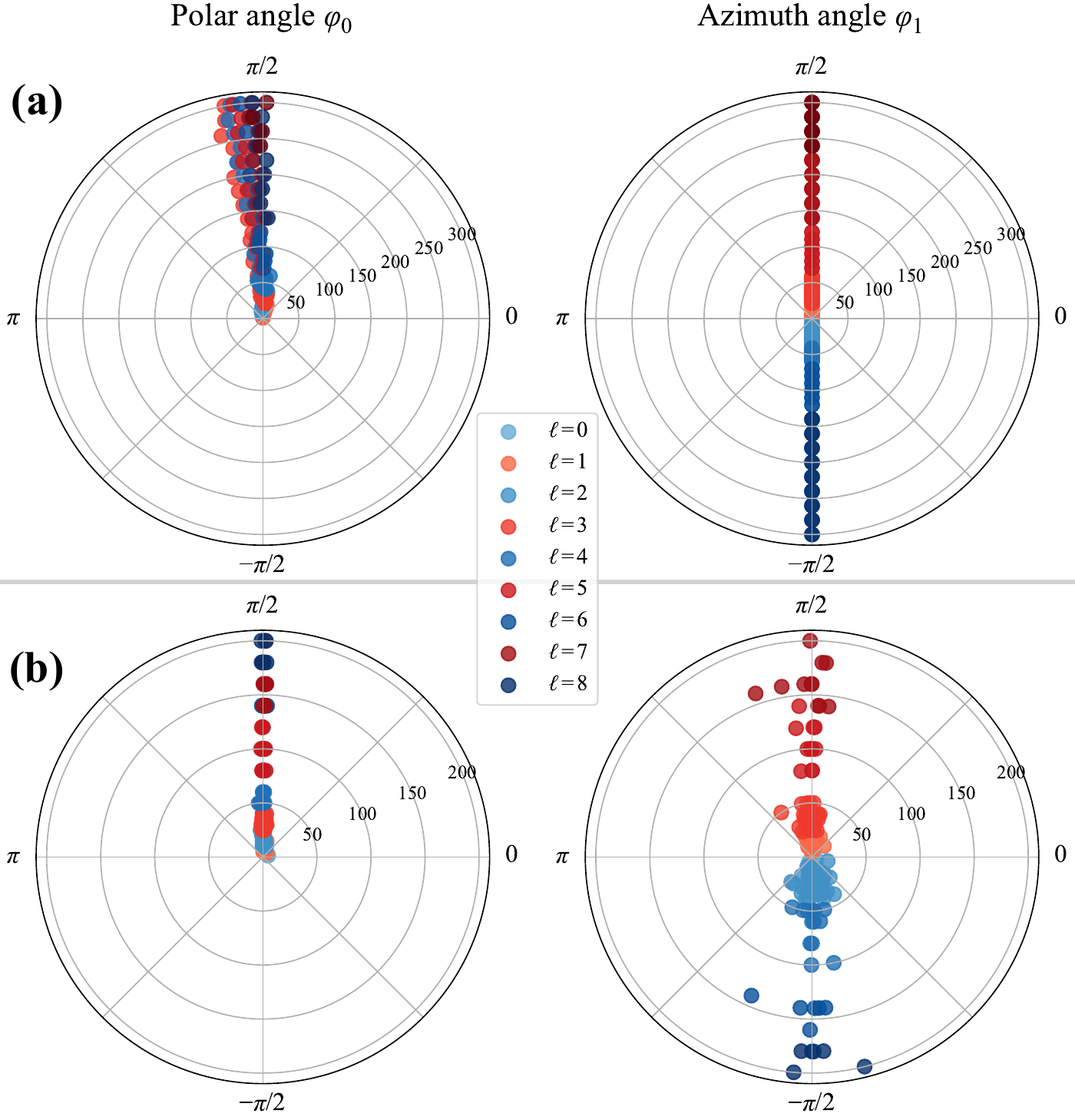}
    \caption{{Distribution of optimal qubit projection angles.}
    Optimized angles for the $p=1$ seed {(a)} and the best-$p$ protocol {(b)}
[cf.\ Fig.~\ref{fig:result}(a)].
    The optima correspond to Bloch vectors near the equator, consistent with the large-$|\alpha|$
    JC half-revival ``attractor''; for $\alpha=\sqrt{N}\in\mathbb{R}^+$ this yields
    $(\varphi_0,\varphi_1)\approx(\pi/2,\pm\pi/2)$ up to conventions.}
    \label{fig:p1_qubit_angle}
\end{figure}

\section{Rugged landscape and transfer initialization}
At revival times, the qubit approaches an equatorial Bloch state,
i.e., an equal-weight superposition of $\ket{g}$ and $\ket{e}$ with
$\langle\sigma_z\rangle\simeq 0$~\cite{GeaBanacloche1990,gea1991atom}.
Starting from $\ket{\Psi_0}=\ket{e}\ket{\alpha}$ with $\alpha=\sqrt{N}$, we write
\begin{eqnarray}
\ket{e}&=&\frac{1}{\sqrt{2}}\bigl(\ket{D_+(0)}-\ket{D_-(0)}\bigr),\\
\ket{D_\pm(0)}&=&\frac{1}{\sqrt{2}}\bigl(\ket{g}\pm\ket{e}\bigr),
\end{eqnarray}
where $\ket{D_\pm}$ are eigenstates of the effective semiclassical Hamiltonian.
In the large-$N$ limit, the two branches evolve approximately independently,
acquiring opposite dynamical phases,
\begin{equation}
\ket{D_\pm(t)}\approx \frac{1}{\sqrt{2}}
\bigl(\ket{g}\pm e^{\mp i\Omega t/(2\sqrt{N})}\ket{e}\bigr).
\end{equation}

At the revival instants $\Omega t=(2\ell+1)\pi\sqrt{N}$, the phase factor
$e^{\mp i\Omega t/(2\sqrt{N})}=e^{\mp i(2\ell+1)\pi/2}=\mp i(-1)^{\ell}$,
so the two branches converge to the same equatorial qubit state,
\begin{equation}
\ket{D_+(t)}\approx \ket{D_-(t)}\equiv
\frac{1}{\sqrt{2}}\bigl(\ket{g}-i(-1)^{\ell}\ket{e}\bigr).
\end{equation}
Thus qubit--cavity entanglement is strongly suppressed and the qubit is nearly pure,
consistent with the optimized projection angles in Fig.~\ref{fig:p1_qubit_angle}.

Optimizing photon-number state preparation amounts to minimizing the loss
$\mathcal{L}_{\bm{\vartheta}}$ over $\bm{\vartheta}=\{\bm{\tau},\bm{\beta},\bm{\varphi}\}$.
The landscape is highly non-convex even in low-dimensional slices: for $p=1$ with fixed
$\mathrm{Im}(\beta_1)$, no projection, and target $N=10$ ($\ell=2$), the reduced loss
$\mathcal{L}_{\bm{\vartheta}}$ shows a dense, quasi-periodic
multi-extremal structure over $\tau_1\in[48,53]$ and $\mathrm{Re}(\beta_1)\in[-0.7,0]$
(Fig.~\ref{fig:nonconvex}).
This structure stems from interference between the two JC dynamical branches in
Eq.~\eqref{eq:Fmm}, readily trapping gradient-based or naive global searches.

These features motivate a transfer-initialized strategy based on the empirically and
analytically established \emph{parameter concentration} phenomenon, where near-optimal
parameters cluster and can be transferred across related instances or circuit depths
(e.g., the concentrated optimal projection angles in Fig.~\ref{fig:p1_qubit_angle})
~\cite{Akshay2021ParamConcentration,Zhou2020QAOA_PRX}.
Accordingly, we adopt a transfer-initialized global-to-local optimizer, described next.

\begin{figure}[!t]
  \centering
  \includegraphics[width=1\linewidth]{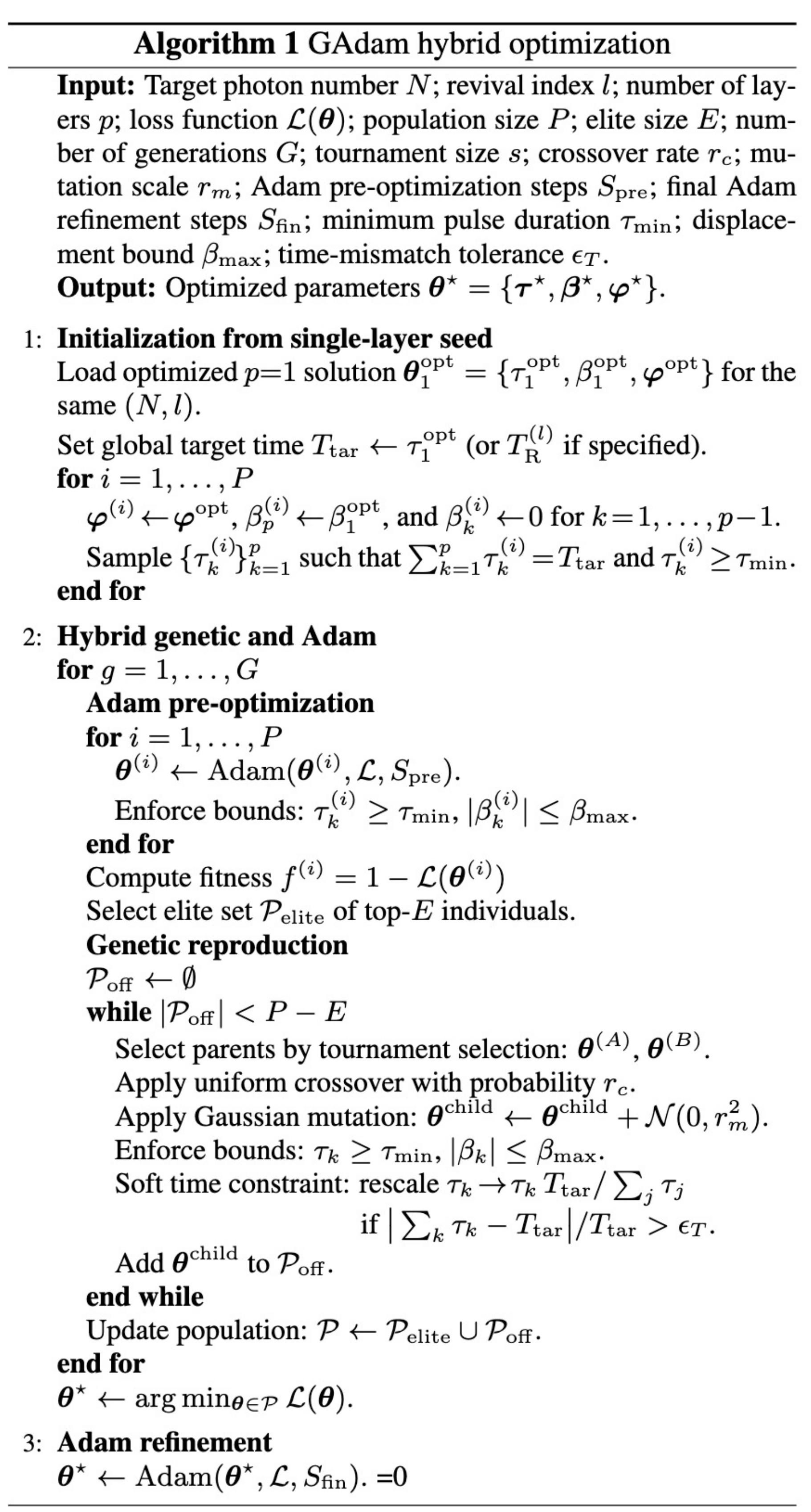}
  \caption{\textbf{GAdam optimization workflow.} }
  \label{alg:gadam}
\end{figure}

\section{Hybrid optimization: GAdam with transfer initialization}
We optimize the $3p+2$ control parameters in Eq.~(\ref{eq:theta}) using a hybrid
GAdam routine (Fig.~\ref{alg:gadam}), which combines global stochastic search with
local Adam refinement to navigate the highly non-convex, multi-minima landscape.

\paragraph*{Transfer initialization.}
As described in Step~1 of Fig.~\ref{alg:gadam}, the optimization is seeded by
a transfer-initialization procedure based on an optimized single-layer ($p=1$)
solution at the same target photon-number $N$ and revival index $\ell$.
The optimized parameters
$\boldsymbol{\theta}_1^{\mathrm{opt}}=\{\tau_1^{\mathrm{opt}},\beta_1^{\mathrm{opt}},\boldsymbol{\varphi}^{\mathrm{opt}}\}$
define a physically meaningful global timing scale.
This total evolution time, denoted $T_{\mathrm{tar}}$, is fixed either to
$\tau_1^{\mathrm{opt}}$ or to the revival-based estimate $T_{\mathrm{R}}^{(\ell)}$.
Initial multi-pulse candidates are generated by randomly partitioning
$T_{\mathrm{tar}}$ into $p$ segments $\{\tau_k\}$ subject to
$\tau_k\ge \tau_{\min}$, while inheriting the optimized final displacement
$\beta_p^{\mathrm{opt}}$ and qubit measurement basis
$\boldsymbol{\varphi}^{\mathrm{opt}}$.
All intermediate displacements are initialized to zero.
This construction places the initial population close to a physically relevant
manifold, avoiding inefficient exploration of low-fidelity regions.

\paragraph*{Hybrid genetic and Adam search.}
The optimization then proceeds iteratively over $G$ generations (Step~2 of
Fig.~\ref{alg:gadam}).
Within each generation, every individual first undergoes a short Adam
pre-optimization using the loss function $\mathcal{L}(\boldsymbol{\theta})$,
which improves the overall population quality before genetic selection.
After enforcing hard bounds on pulse durations and displacement amplitudes, the
fitness of each individual is evaluated as
$f=1-\mathcal{L}(\boldsymbol{\theta})$.
An elite subset of the top-$E$ individuals is retained unchanged.

New candidates are generated through tournament-based parent selection,
uniform crossover of control parameters, and Gaussian mutation.
A soft global timing constraint is imposed: if the total duration deviates from
$T_{\mathrm{tar}}$ beyond a relative tolerance $\epsilon_T$, all $\tau_k$ are
rescaled proportionally to restore the target total time.
This allows the optimizer to freely redistribute pulse durations while
preserving the global interference timescale set by the JC revival structure.
The next generation is formed by combining the elite set with the newly
generated offspring.

\paragraph*{Final refinement.}
After $G$ generations, the individual with the minimal loss is selected and
subjected to a final, high-precision Adam refinement (Step~3 of
Fig.~\ref{alg:gadam}).
The resulting parameters
$\boldsymbol{\theta}^\star=\{\boldsymbol{\tau}^\star,\boldsymbol{\beta}^\star,\boldsymbol{\varphi}^\star\}$
define the optimized composite-pulse sequence used in the performance and
robustness analyses.

\bibliography{mybib}

\end{document}